# A galaxy at a redshift $z$=6.96

Masanori Iye[1,2,3], Kazuaki Ota[2], Nobunari Kashikawa[1], Hisanori Furusawa[4], Tetsuya Hashimoto[2], Takashi Hattori[4], Yuichi Matsuda[5], Tomoki Morokuma[6], Masami Ouchi[7] & Kazuhiro Shimasaku[2]

[1]National Astronomical Observatory, 2-21-1 Osawa, Mitaka, Tokyo 181-8588, Japan. [2]Department of Astronomy, University of Tokyo, Hongo, Bunkyo-ku, Tokyo 113-0033, Japan. [3]The Graduate University for Advanced Studies, Mitaka, Tokyo 181-8588, Japan. [4]Subaru Telescope, 650 North A'ohoku Place, Hilo, Hawaii 96720, USA. [5]Department of Astronomy, Graduate School of Science, Kyoto University, Kyoto 606-8502, Japan. [6]The Institute of Astronomy, University of Tokyo, Mitaka, Tokyo 181-0015, Japan. [7]Space Telescope Science Institute, 3700 San Martin Drive, Baltimore, Maryland 21218, USA.

**When galaxy formation started in the history of the Universe remains unclear. Studies of the cosmic microwave background indicate that the Universe, after initial cooling (following the Big Bang), was reheated and reionized by hot stars in newborn galaxies at a redshift in the range 6<$z$<14 (ref. 1). Though several candidate galaxies at redshift z>7 have been identified photometrically[2,3], galaxies with spectroscopically confirmed redshifts have been confined to $z$<6.6 (refs 4–8). Here we report a spectroscopic redshift of $z$=6.96 (corresponding to just 750 Myr after the Big Bang) for a galaxy whose spectrum clearly shows Lyman-α emission at 9,682 Å, indicating active star formation at a rate of ~$10 M_{\odot}$ $yr^{-1}$, where $M_{\odot}$ is the mass of the Sun. This demonstrates that galaxy formation was under way when the Universe was only ~6 per cent of its present age. The number density of galaxies at $z$≈7 seems to be only 18–36 per cent of the density at $z$=6.6.**

Narrowband filter surveys searching for redshifted Lyman-α emitting galaxies (LAEs) through dark atmospheric windows in which the foreground telluric OH band emission is not prohibitively bright have been successful in isolating galaxies at high redshift[4,5]. Some of the discovered LAEs are gravitationally lensed objects behind clusters of galaxies. To obtain unbiased information about the population of LAEs, systematic surveys at various redshifts have been performed, for example, in the Subaru Deep Field (SDF) [6-8]. Such studies have been limited to a redshift range of $z$<6.6. New attempts are in progress[9-12] to search for galaxy populations at 7<$z$<10, using deep narrowband imaging observations with near-infrared cameras, aiming for high sensitivity despite the limited survey volume. We have taken another approach to survey $z$=7 LAEs, specifically using the last OH window accessible with a wide-field charge-coupled device (CCD) camera.

We developed a narrowband filter NB973 for the Subaru Suprime-Cam[13,14], with a transmission bandwidth of 200 Å centred at 9,755 Å. To investigate the population of $z$=7.0 LAEs, we conducted an NB973 imaging survey with 15 h exposure (NB973<24.9 mag in 5$\sigma$) on the photometric nights of 16 and 17 March 2005. An effective area of 876 arcmin$^2$ and co-moving depth of 58 Mpc, corresponding to $z$=6.94–7.11, led to a total probed volume of 3.2×10$^5$ Mpc$^3$. (Throughout, we adopt a cosmology with $\Omega_M$=0.3, $\Omega_\Lambda$=0.7, and $H_0$=70 km s$^{-1}$ Mpc$^{-1}$ and refer to magnitudes in the AB system measured with a 2″ aperture unless otherwise specified.) We measured the NB973 fluxes of all detected objects and also measured their broadband fluxes using images previously obtained by the SDF project[8]. The colours of objects were derived from these measurements and compared to the expected colours of a $z$=7.0 LAE, computed using a stellar population synthesis model[15] and an intergalactic absorption model[16]. Of the 41,533 objects detected in NB973, we found only two objects with significant NB973 flux that were not detected in any of the broadband images blueward of NB973. We considered these to be $z$=7 LAE candidates (hereafter referred to as IOK-1, the brighter of the two in NB973, and IOK-2). Their images and photometric properties, as well as the candidate selection criteria, are shown in Fig. 1 and Table 1.

To confirm whether these candidates are real LAEs, we carried out follow-up spectroscopy on 14 and 15 May, 1 June 2005, and 24 April 2006, using the Faint Object Camera and Spectrograph (FOCAS)[17]. Total exposures were 8.5 h and 3.0 h for IOK-1 and IOK-2, respectively. Figure 2 shows the spectrum of IOK-1 with OH skylines subtracted. We identified this object as a LAE at $z$=6.96, which makes it, to our knowledge, the most distant galaxy ever spectroscopically confirmed. Our discovery of this galaxy provides direct evidence that at least one galaxy already existed only ~750 million years after the Big Bang.

Three-hour integration on IOK-2 did not reveal significant spectral features, and our photometric detection could have been due to noise. There are 10$^6$ independent 2″ apertures in the 876 arcmin$^2$ field, and the probability of noise exceeding 5$\sigma$ for a gaussian distribution is 3×10$^{-7}$, so that the expected number of 5$\sigma$ noise events is 0.3. Another possible origin for IOK-2 is a transient object such as a supernova or active galactic nucleus that happened to brighten in March 2005, but was inconspicuous in broadband images taken 1–2 years earlier. Counts of variable objects in i' band images from the SDF taken over 1–2 years indicate that the expected number of variable objects having brightness amplifications large enough to exceed our detection limit (NB973=24.9) is at most about three. Those transient objects, if any existed, would not show conspicuous emission line features in the NB973 waveband.

We note, however, that even for the brighter IOK-1, it was necessary to combine the eleven 30-min exposures with the highest signal-to-noise ratio under the best sky conditions selected from the 17 spectra taken to detect the $z$=6.96 Lyman-α emission shown in Fig. 2 clearly. Because our spectral detection limit for 3 h did not reach the depth required to deny the presence of weak Lyman α emission, we retain IOK-2 as another possible but unconfirmed LAE. Therefore, we conclude that the detected number of LAEs at $z$=7.0 in our survey volume is at least one, and possibly could be two at most.

The observed LAE number density is expected to decline beyond the redshift at which the reionization had completed, because the increasing fraction of intergalactic neutral hydrogen absorbs the Lyman-α photons from young galaxies. However, no significant decline of LAE number densities has been yet observed for 3<$z$<6.6 (ref. 26). Figure 3 shows the volume densities of the LAE numbers, Lyman-α line luminosity, and star formation rate at three redshifts[6,17,18] down to our detection limit. These physical parameters are useful for tracing the properties of LAEs in an environment with an increasing fraction of neutral hydrogen in the intergalactic medium (IGM). Errors shown in the figure include the poissonian error associated with small-number statistics and the cosmic variance uncertainty in galaxy number density due to the nonuniform distribution of galaxies that reflect the large-scale structure of the Universe. We estimated the amount of expected cosmic variance using an analytic cold dark matter model[19] assuming a bias parameter of LAEs, $b$=3.4±1.8 (ref. 20), and taking SDF survey volumes and LAE number densities at $z$=6.6 (ref. 7) and 7.0. The resultant cosmic variance of $z$=6.6–7.0 LAEs in the SDF down to our detection limit is at most 25%, which is less than the poissonian error described below.

The $z$=6.6 LAE survey[7] of the same field (SDF), but with a smaller volume (2.2×10$^5$ Mpc$^3$), photometrically yielded four bright LAEs down to a comparable luminosity limit. Assuming that the evolution of the LAE population is negligible, we should have $6^{+3.9}_{-2.8}$ LAEs within our survey volume at $z$=7.0. However, our finding of only one (or at most two) LAE(s) at $z$=7.0 suggests that the LAE number density is only 18% (or 36%) of that at $z$=6.6. The confidence level of the difference in the number density between $z$=7.0 and $z$=6.6 is 93% (or 85%), taking the poissonian error for small-number statistics[21]. The observed variation may have been due to the evolutionary change in the LAE properties. However, there is no quantitative way to assess this effect reliably for the moment.

The decrease in the LAE number density at $z=7.0$ compared with $z=6.6$, if confirmed in other fields and for the fainter end of the population, could reflect a deficit in Lyman-α photons owing to attenuation by the neutral IGM. A similar decrease in the number density of LAEs was also recently found over the redshift range $5.7<z<6.6$ (ref. 22). The neutral hydrogen fraction of the IGM, $x_{HI}$, at $z=7.0$ is estimated to be about 0.63–0.83, applying the ratio of the luminosity density at $z=7.0$ to that at $z=6.6$ to the Lyman-α line attenuation model[23] and assuming complete reionization at $z=6.6$. Note that our result does not contradict the Gunn–Peterson test[24] (diagnostics of intergalactic Lyman-α absorptions on high redshift quasars) that infers the completion of reionization at $z\approx6.0$ (ref. 25). This is because the LAE number or luminosity density is sensitive to $x_{HI} \geq 0.1$ (ref. 26), whereas the quasar Gunn–Peterson trough is sensitive to $x_{HI} \approx 0.01$ (ref. 25). Therefore, it is natural that the LAE density should not change much at $z<6.6$, when the quasar Gunn–Peterson troughs show significant attenuation of Lyman-α photons.

Note, however, that the present survey is not sufficiently deep to sample the entire LAE population ($5\sigma$ detection limit $\geq 9.9 \times 10^{42}$ erg s$^{-1}$). Follow-up imaging and spectroscopic surveys in the SDF are planned, to probe fainter $z=7.0$ LAEs by using red-sensitive CCDs to be updated on Suprime-Cam. Independent surveys of other fields are essential for investigating cosmic variance more accurately and for studying field-to-field variations in reionization status.

**Acknowledgements** This work is based on data collected with the Subaru Telescope, operated by the National Astronomical Observatory of Japan and was partly supported by KAKENHI. We thank the SDF team and the staff at the Subaru Observatory for providing the data and assisting us with observations and data reduction. K.O. acknowledges a fellowship from the Japan Society for the Promotion of Science.

**Author Contributions** M.I., the principal investigator, proposed this project, designed the NB973 filter, and led the observations. K.O. made all the observations, including the slit mask design for the spectroscopy, carried out all the data reduction and analysis, and produced the figures. M.I. and K.O. jointly wrote the paper. N.K. contributed to observations, data analysis, and discussion. H.F. and Tk.H. helped with the Suprime-Cam imaging and FOCAS spectroscopy, respectively, as instrument support astronomers. K.S. and M.O. commented on the manuscript. T.M. investigated the possibility of the unconfirmed candidate being a variable object. Y.M., K.O. and Tt.H. measured the transmission curve of the NB973 filter.

**Author Information** Reprints and permissions information is available at npg.nature.com/reprintsandpermissions. The authors declare no competing financial interests. Correspondence and requests for materials should be addressed to M.I. (iye@optik.mtk.nao.ac.jp).

**Table 1 Properties of $z\approx 7$ Lyman-$\alpha$ emitter candidates**

| ID | Position(j2000) | i '(mag) | z' (mag) | NB973 (2″) | NB973 (total) | $L$(Ly$\alpha$) ($10^{43}$ erg s$^{-1}$) | SFR(Ly$\alpha$)$\varepsilon$ ($M_\bullet$ yr$^{-1}$) |
|---|---|---|---|---|---|---|---|
| IOK-1 | $\alpha$=13$^h$ 23$^m$ 59.$^s$8 $\delta$=+27° 24′ 55.″8 | >27.84 | >27.04 | 24.60 | 24.40 | 1.1±0.2 | 10±2 |
| IOK-2 | $\alpha$=13$^h$ 25$^m$ 32.$^s$9 $\delta$=+27° 31′ 44.″7 | >27.84 | >27.04 | 25.51 | 24.74 | <1.1±0.2 | <10±2 |

Magnitudes (SDSS i', z' filters and NB973) are in the AB system, measured with a 2″ aperture. The NB973 total magnitudes integrated over the entire image are also given. The Lyman-$\alpha$ luminosity $L$(Ly$\alpha$) of IOK-1 was derived from spectroscopy, while the upper limit of $L$(Ly$\alpha$) for IOK-2 was evaluated using NB973 total magnitude by assuming that 68% of the NB973 flux is in the Lyman-$\alpha$ line, on the basis of simulations. Luminosities were then converted into corresponding Lyman-$\alpha$ star-formation rate SFR(Ly$\alpha$) using the relation derived from Kennicutt's equation[27] with case B recombination theory. 1$\sigma$ errors are also given for $L$(Ly$\alpha$) and SFR(Ly$\alpha$). IOK-1 was spectroscopically identified as a $z$=6.96 Lyman-$\alpha$ emitter. For IOK-2, the photometric estimation of Lyman-$\alpha$ luminosity should be regarded as an upper limit because we were unable to make a spectroscopic confirmation.

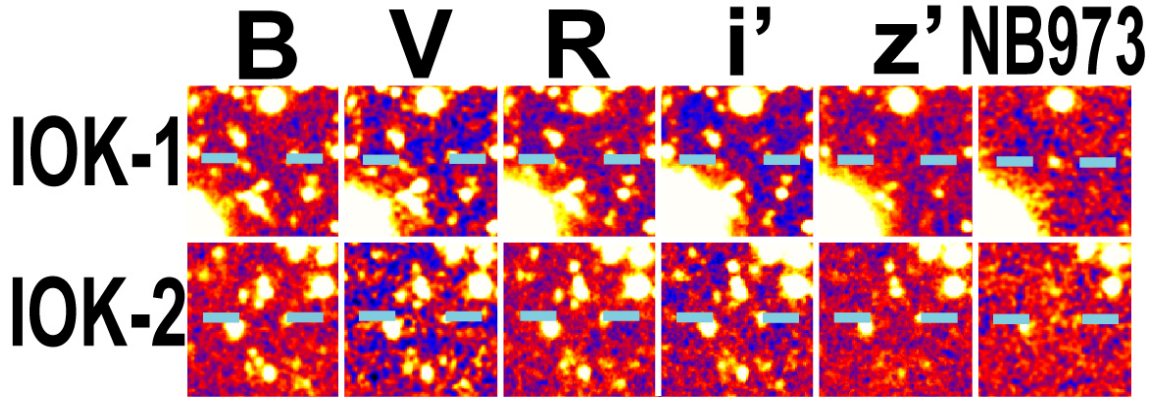

**Figure 1 Multi-waveband 20″×20″ images of the $z$=6.96 Lyman $\alpha$ emitter IOK-1 and the unidentified candidate IOK-2.** Deep broadband images (Kron–Cousin B, V, R bands, and SDSS i' and z' bands, with limiting magnitudes 28.45, 27.74, 27.80, 27.43 and 26.62 in 3$\sigma$, respectively) were taken during 2002–2004 (ref. 8). Dithered NB973 exposures (15–30 min each; 15 h total) were combined to remove fringing produced by foreground sky OH emission lines and yielded a 2″ aperture limiting magnitude of NB973=24.9 (5$\sigma$). All the images were convolved to a common seeing size of 0.″98. Of the 41,533 objects detected down to NB973=24.9 (total magnitude in this case), we selected those satisfying the following criteria: (1) B, V, R, i', z'<3$\sigma$, for $z$=7.0 LAE candidates IOK-1 and IOK-2, and (2) B, V, R <3$\sigma$, i'–z' >1.3, z'-NB973 >1.0, for

z=6.4–7.0 LAE candidate IOK-3. These criteria were determined by model calculations[15,16]. The i'–z' colour eliminates interlopers such as lower-$z$ [O II], [O III] line-emitting galaxies while the colour z'-NB973 picks up objects with NB973 flux excess with respect to the z' continuum flux. In addition, null (<3$\sigma$) detection in B, V and R (or B, V, R, i' and z') total magnitudes are required since continuum fluxes blueward of the Lyman-$\alpha$ line should be mostly (or completely) absorbed by the neutral IGM. The criterion-(2) object, IOK-3, was identified as a $z$=6.6 LAE by an independent spectroscopic survey[22] and the possibility of a $z$=7.0 LAE was excluded.

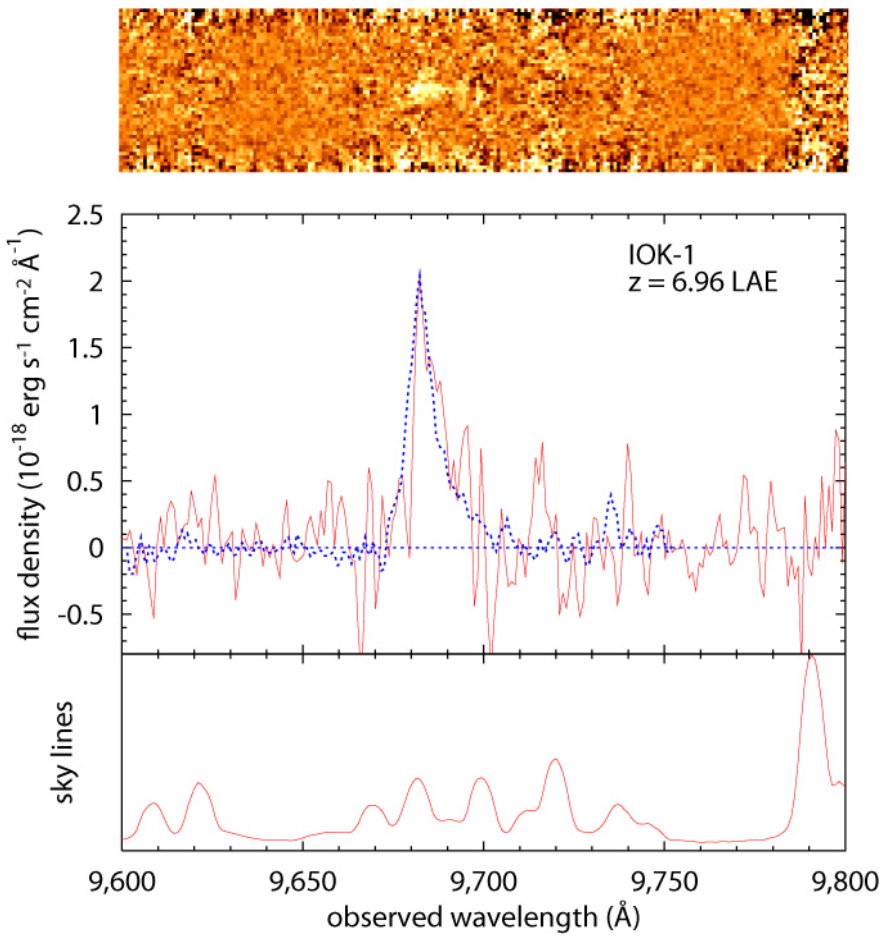

**Figure 2 Combined spectrum of $z$=6.96 galaxy, IOK-1.** The bottom panel shows the removed OH sky emission lines. An echelle grism (175 lines per millimetre, resolution≈ 1,600 ) with z' filter and 0.8" slit was used to obtain 11 spectra of 30-min exposure each, dithered along the slit by ±1". The Lyman-$\alpha$ emission peak is located at 9,682 Å. This feature was confirmed to follow the dithered shifts along the slit. The Lyman-$\alpha$ emission

has a total flux of $2.0\times10^{-17}$ erg s$^{-1}$ cm$^{-2}$, consistent with the estimate from the NB973 image ($2.7\times10^{-17}$ erg s$^{-1}$ cm$^{-2}$), and a full width at half-maximum of 13 Å. The Lyman-α luminosity is $L(Ly\alpha)=1.1\times10^{43}$ erg s$^{-1}$, corresponding to $SFR(Ly\alpha)=10$ $M_\odot$ yr$^{-1}$. The emission feature is significant at the $5.5\sigma$ level. The asymmetric emission line profile matches the composite template line profile (dashed line) produced from 12 Lyman-α emitters at $z=6.6$ (ref. 22) normalized and shifted to $z=6.96$. This emission cannot be unresolved $z=1.6$ [O II] doublet lines with 7.0 Å separation (3,726 Å, 3,729 Å at rest frame), as finer sky lines (about 5.1–6.2 Å) are clearly resolved. Similarly, the line is not Hβ, [O III] 4,959 Å, [O III] 5,007 Å, Hα, [S II] 6,717 Å or [S II] 6,731 Å, because our spectrum shows none of the other lines that should accompany any of these lines in the observed wavelength range. The possibility that OH lines mask other lines was carefully examined for each case and ruled out. There appears to be an extremely weak emission at around $z=7.02$ in the 3 h integration spectrum of IOK-2, but this needs further integration for confirmation.

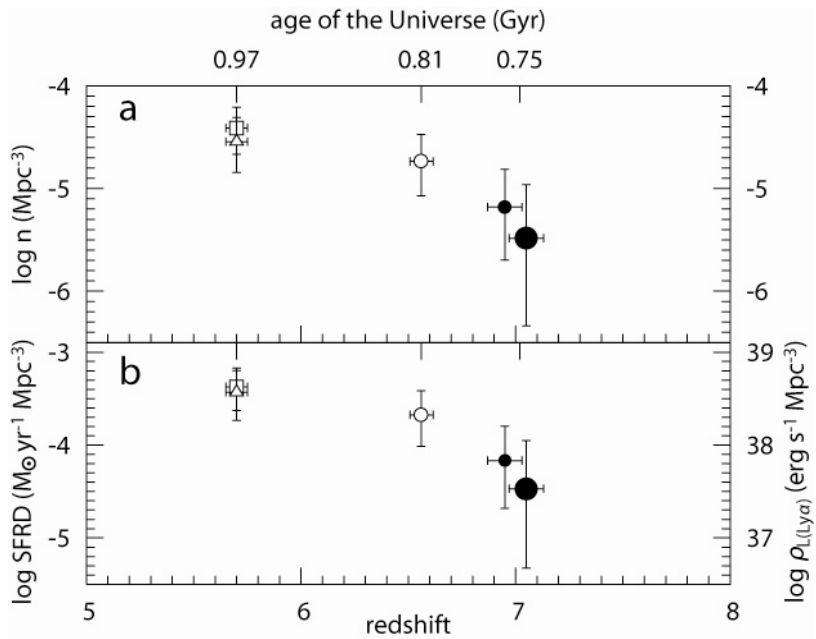

**Figure 3 Decline of the number density of LAEs between 6.6<$z$<7.0.** The number density $n$ (**a**), luminosity density, $\rho_{L(Ly\alpha)}$, and SFRD (**b**) derived from Lyman-α line fluxes detected in several Suprime-Cam LAE surveys are plotted for the three epochs of $z=5.7$ (square[6] and triangle[18]: other survey fields), 6.6 (open circle[7]: also SDF), and 7.0 (filled circles: our result) down to our detection limit $L(Ly\alpha)=9.9\times10^{42}$ erg s$^{-1}$ calculated

from NB973=24.9 (5$\sigma$). The volumes probed by these surveys are of a similar order of magnitude (~2–3×10$^5$ Mpc$^3$). The large filled circle at $z$=7.0 represents the values derived for the single LAE discovered in the present study, and the small filled circle those for two LAEs, including the unconfirmed candidate IOK-2. All the vertical error bars in both panels include the errors from both Poisson statistics and cosmic variance. The horizontal error bars indicate the surveyed redshift ranges. Note that large and small filled circles are shifted slightly horizontally from $z$=7.0 to make them easier to see.